\documentclass[sn-mathphys]{sn-jnl}

\jyear{2022}

\usepackage{soul}

\hypersetup{dvips, colorlinks=true, linkcolor=black, citecolor=black, filecolor=blue, urlcolor=black}

\newcommand{\bL}{\mathbf{L}}
\newcommand{\bK}{\mathbf{K}}
\newcommand{\bD}{\mathbf{D}}
\newcommand{\bv}{\mathbf{v}}
\newcommand{\bn}{\mathbf{n}}
\newcommand{\mH}{\mathcal{H}}
\renewcommand{\p}{\partial}
\newcommand{\us}{\underline{s}}
\newcommand{\bsigma}{\boldsymbol{\sigma}}
\newcommand{\btheta}{\boldsymbol{\theta}}
\newcommand{\bomega}{\boldsymbol{\omega}}

\newcommand{\re}{\mathrm{e}}
\newcommand{\rI}{\mathrm{I}}

\newcommand{\eps}{\epsilon}
\newcommand{\rd}{\mathrm{d}}
\newcommand{\const}{\mathrm{const.}}

\begin{document}

\title[Milankovitch equations with spinors]{Milankovitch equations with spinors}

\author[1]{\fnm{Barnab\'{a}s} \sur{Deme}}\email{deme@iap.fr}
\author[1]{\fnm{Jean-Baptiste} \sur{Fouvry}}\email{fouvry@iap.fr}

\affil[1]{CNRS and Sorbonne Universit{\'e}, UMR 7095, Institut d'Astrophysique de Paris, 98 bis Boulevard Arago, F-75014 Paris, France}

\abstract{We investigate the use of spinors to describe the secular evolution
of quasi-Keplerian systems. Evaluating their Poisson brackets, we show that the components of a properly-chosen spinor are canonical variables.
We illustrate this formalism with a satellite's motion around an oblate body.}


\maketitle

\section{Introduction}
\label{sec1}

The three-body problem is one of the oldest open problems in astronomy \citep{valtonen06}. Although it can be applied as a good model to many astrophysical phenomena ranging from the dynamics of minor planets~\citep{kozai62} and satellites~\citep{lidov62} to that of supermassive black holes~\citep{naoz19,alexander17},
it cannot be generically solved.\footnote{There exists a solution in the form of an infinite series~\citep{sundman}, but it is of no practical use due to its slow convergence~\citep{diacu97}.} It is non-integrable, i.e., chaotic~\citep{masoliver2010}. 

There exist a couple of approaches to predict the future of triple systems: (i) one directly integrates the equations of motion numerically~\citep{heggie03}; (ii) one derives the probability distributions of the orbital elements as a result of the chaotic three-body evolution~\citep[e.g.,][]{stone2019}; (iii) one uses approximations~\citep{valtonen06}. A frequently used approximation is the \textit{restricted} three-body problem in which one of the three bodies has zero mass~\citep{szebehely1967}. Another possibility is the \textit{hierarchical} problem in which a tight binary is perturbed by a more massive object~\citep{naoz2016}. The hierarchical three-body problem puts the focus on the system's evolution on secular timescales,
i.e., timescales much longer than the orbital period. It is achieved by averaging the system's Hamiltonian, for example using the von Zeipel transformation~\citep{ito2019}. The secular evolution equations for the orbit can then be obtained from the averaged Hamiltonian in different ways. 

A common method is Lagrange's (non-canonical) planetary equations of motion
which use the standard orbital elements~\citep[e.g.,][]{murraydermott99}.
However, the associated equations suffer from vanishing denominators
if the eccentricity is either zero or unity, or if the inclination is zero~\citep{valtonen06}.

In order to cure the aforementioned problems, one can introduce another set of equations, named after Milankovitch~\citep{milankovitch39}. More precisely, one defines
$\bL$ as the angular momentum vector and $\bK$ (the pericentre vector) as the vector which points to the orbit's pericentre
and whose magnitude is equal to the angular momentum.
The (non-canonical) evolution equations then read
\begin{subequations}
\label{eq:LKdot}
\begin{align}
    \dot{L}_a {} & =\{L_a, \mH \}=\{L_a,L_b\}\frac{\p \mH}{\p L_b} \; +\{L_a,K_b\}\frac{\p \mH}{\p K_b},
\\
    \dot{K}_a {} & =\{K_a, \mH \}=\{K_a,L_b\}\frac{\p \mH }{\p L_b}+\{K_a,K_b\}\frac{\p \mH}{\p K_b},
\end{align}
\end{subequations}
where Einstein summation is implied for the Latin indices, which run from 1 to 3.
Here, $\mH$ is the system's Hamiltonian and $\{\cdot,\cdot\}$ is the Poisson bracket
\begin{equation}
    \{ f(q,p), g(q,p) \} = \frac{\p f}{\p q} \frac{\p g}{\p p} - \frac{\p f}{\p p}\frac{\p g}{\p q},
\end{equation}
with $f$ and $g$ arbitrary functions of ${ (q,p) }$, the canonical coordinates and momenta~\citep{goldstein02}.  These equations are not limited to the three-body problem, i.e., $\mH$ is arbitrary.\footnote{The standard version of Milankovitch equations uses the Runge--Lenz--Laplace or eccentricity vector instead of our pericentre vector~\citep[e.g.,][]{tremaine2009,Rosengren2014}. Some other works~\citep[e.g.,][]{fouvry2022}, work with the Klein variables~\citep{klein1924}. The reason for our choice is clarified in Sec.~\ref{sec:general}.}
For a bound orbit, the Poisson brackets of these vectors are
\begin{subequations}
\label{eq:Poisson_LK}
\begin{align}
    \{L_a,L_b\} & = \;\;\,\, \eps_{abc} L_c,  \label{eq:poisson_LL}
    \\
    \{K_a,L_b\} & = \;\;\,\, \eps_{abc} K_c,  \label{eq:poisson_KL}
    \\
    \{K_a,K_b\} & = - \eps_{abc} L_c, \label{eq:poisson_KK}
\end{align}
\end{subequations}
with $\eps_{abc}$ the Levi--Civita anti-symmetric tensor.
We check these in Appendix~\ref{app:vectors}
using the canonical Delaunay variables.
Inserting these expressions into Eqs.~\eqref{eq:LKdot} yields
\begin{subequations}
\label{eq:milankovitch_vec}
\begin{align}
    \dot{L}_a {} & = \eps_{abc} L_c \frac{\p \mH}{\p L_b} \;- \eps_{abc} K_b \frac{\p \mH}{\p K_c} ,
\\
    \dot{K}_a {} & = \eps_{abc} K_c \frac{\p \mH}{\p L_b} + \eps_{abc} L_b \frac{\p \mH}{\p K_c} .
\end{align}
\end{subequations}
Hereafter we restrict ourselves to secular evolution. An essential feature of it is that the semi-major axis is constant due to the orbit-averaging over the (fast) mean anomaly~\citep{morbidelli2002}. For a time-independent averaged Hamiltonian, i.e., when the total energy is constant as well, we have two conserved quantities. These are manifested in the first integrals of motion of Eqs.~\eqref{eq:milankovitch_vec}
\begin{subequations}
\label{eq:constraints_LK}
\begin{align}
   \|\bL\| - \|\bK\| {} & = \const ,
\\
   \bL \cdot \bK {} & = \const ,
\end{align}
\end{subequations}
with ``$\cdot$'' the usual scalar product. The associated constants are set to zero because $\bL$ and $\bK$ have the same magnitudes and are perpendicular to each other, by definition.
These two constraints reduce the number of degrees of freedom from 6 to 4. 

Contrary to the standard vector algebra just presented, Hestenes (1983)~\cite[][H83 hereafter]{hestenes83} suggested an alternative formulation for celestial mechanics, namely geometric algebra. In doing so, H83 derives the equations of motion for a spinor instead of  a pair of vectors. We now revisit this result and rewrite the associated equations of motion using the Poisson brackets of spinors. The letter is structured as follows. Section~\ref{sec:general} introduces spinors and expresses the equations of motion with them. Section~\ref{sec:oblate} applies the spinorial equations to the pedagogical case of the secular dynamics around an oblate planet. Section~\ref{discussion} summarises the benefits and disadvantages of the spinorial formalism.

\section{The spinorial Milankovitch equations}
\label{sec:general}

First, let us take three orthonormal unit vectors $\bsigma_a \in \mathbb{R}^{3} $
with $a=1,2,3$. Let us then define the geometric product between two of them as
\begin{equation}
    \bsigma_a \bsigma_b = \bsigma_a \cdot \bsigma_b + \bsigma_a \wedge \bsigma_b,
\end{equation}
where ``$\cdot$'' is again the usual (symmetric) dot product giving a scalar, while ``$\wedge$'' is an antisymmetric product giving a bivector. From this, it follows that
\begin{equation}\label{eq:clifford}
    \bsigma_a \bsigma_b = - \bsigma_b \bsigma_a
\end{equation}
for $a\neq b$ and
\begin{equation}\label{eq:clifford2}
    \bsigma_1 \bsigma_1 = \bsigma_2 \bsigma_2 = \bsigma_3 \bsigma_3 = 1 ,
\end{equation}
where $1$ is a scalar and the geometric product is not denoted explicitly.\footnote{The notation $\bsigma_a$ is motivated by the fact that the Pauli matrices, which satisfy exactly the same algebra, are usually denoted in this way, too.} Using the associativity of the geometric product, one can easily show that 
\begin{equation}\label{eq:pauli}
    (\bsigma_1 \bsigma_2 \bsigma_3)^2 = (\bsigma_1 \bsigma_2 \bsigma_3)(\bsigma_1 \bsigma_2 \bsigma_3) = - 1,
\end{equation}
which motivates a similar notation to that of the complex unit,
namely $\bsigma_1 \bsigma_2 \bsigma_3 = \rI$, a trivector.
Now we can define a spinor and its conjugate via
\begin{subequations}
\begin{align}
    \us {} & = x_0 + \mathrm{I} x_a \bsigma_a ,
\\
    \us^\dagger {} & = x_0 - \mathrm{I} x_a \bsigma_a,
\end{align}
\end{subequations}
where $x_{0 \leq \mu \leq 3}$ are real. As illustrated in Appendix~\ref{app:spinor}, spinors can be efficiently used to describe rotations.\footnote{For a detailed introduction into the application of geometric algebra in physics, see~\citep{hestenes71,hestenes02}.}
Along the same line,
Appendix~\ref{app:bivectors} investigates the use of bivectors in the context of Milankovitch equations.

Let us now demand that (i) the norm of the spinor, $\us$, is related to the angular momentum via
\begin{equation}\label{eq:norm}
    \us^\dagger \us = 4 \, \|\bL\| = 4 \, \|\bK\|,
\end{equation}
and that (ii) it rotates the basis vectors $\bsigma_1$ and $\bsigma_3$ to the direction of $\bK$ and $\bL$, respectively (we follow H83 for that convention), i.e.,
\begin{subequations}
\label{eq:rotating_sigma}
\begin{align}
    \bL {} & =\us^\dagger \bsigma_3 \, \us ,
\\
    \bK {} & =\us^\dagger \bsigma_1 \, \us .
\end{align}
\end{subequations}
These equations can be recast as
\begin{subequations}
\begin{align}
    4 L_a \bsigma_a {} & = \left(x_0- \rI x_b \bsigma_b \right) \bsigma_3 \left(x_0 + \rI x_c \bsigma_c \right) ,
    \\
    4 K_a \bsigma_a {} & = \left(x_0 - \rI x_b \bsigma_b \right) \bsigma_1 \left(x_0 + \rI x_c \bsigma_c \right).
\end{align}
\end{subequations}
Matching the prefactors of the $\{ \bsigma_i \}$ on both sides finally yields
\begin{subequations}
\label{eq:spinor_LK}
\begin{align}
    L_1 {} &= \tfrac{1}{2} \big( x_0 x_2 + x_1 x_3 \big) ,
    \\
    L_2 {} &= \tfrac{1}{2} \big(x_2 x_3 - x_0 x_1 \big) ,
    \\
    L_3 {} &= \tfrac{1}{4} \big( x_0^2 - x_1^2 - x_2^2 + x_3^2 \big) ,
     \\
    K_1 {} &= \tfrac{1}{4} \big( x_0^2 + x_1^2 - x_2^2 - x_3^2 \big) ,
    \\
    K_2 {} &= \tfrac{1}{2} \big( x_0 x_3 + x_1 x_2 \big) ,
    \\
    K_3 {} &= \tfrac{1}{2} \big( x_1 x_3 - x_0 x_2 \big) .
\end{align}
\end{subequations}
The action of the spinor on the basis vectors is to rotate and multiply both vectors by $4\|\bL\|$.
Such an operation cannot provide us with two vectors of different magnitudes,
hence our choice in Sec.~\ref{sec1} of using the pericentre vector
rather than the eccentricity one.

We now express the Hamiltonian with the spinor ${ \mH (\us) \!=\! \mH(x_0,x_1,x_2,x_3) }$. Analogously to the vectorial case in Eqs.~\eqref{eq:LKdot}, we have
\begin{equation}\label{eq:xidot}
    \dot{x}_\mu = \{x_\mu, x_\nu\} \p_{x_\nu} \mH,
\end{equation}
where Greek indices run from 0 to 3. The task is to calculate the Poisson brackets of the ${ \{ x_{\mu} \} }$, just like for ${(\bL , \bK)}$ in Eqs.~\eqref{eq:Poisson_LK}. Substituting Eqs.~\eqref{eq:spinor_LK} into the Poisson brackets from Eqs.~\eqref{eq:Poisson_LK} gives us
\begin{subequations}
\label{eq:poissonx}
\begin{align}
    \{x_1,x_2\} & = \{x_3, x_0\} = 1,
    \label{eq:poisson_x1x2}
    \\
    \{x_2,x_3\} & = \{x_0, x_1\} = 0,
    \\
    \{x_0,x_2\} & = \{x_3, x_1\} = 0.
\end{align}
\end{subequations}
Putting these brackets back into Eq.~\eqref{eq:xidot}
finally yields the equations of motion
\begin{subequations}
\label{eq:xdot}
\begin{align}
    \dot{x}_0 {} & = - \p_{x_3} \mH,
    \\
    \dot{x}_1 {} & = \;\,\,\, \p_{x_2} \mH,
    \\
    \dot{x}_2 {} & = - \p_{x_1} \mH,
    \\
    \dot{x}_3 {} & = \;\,\,\, \p_{x_0} \mH.
\end{align}
\end{subequations}
These are the \textit{spinorial} analogues of Milankovitch Eq.~\eqref{eq:milankovitch_vec}. We make two remarks about them. First, the transformations in Eqs.~\eqref{eq:spinor_LK} have the same structure as the KS transformation in the regularisation of the 3-dimensional Kepler problem~\citep{ks65,waldvogel06}.
Here, they are applied to both the angular momentum and pericentre vectors. Second, the set ${ \{ x_{\mu} \} }$ is symplectic, i.e., it obeys the canonical Poisson relations.
Phrased differently, Eqs.~\eqref{eq:xdot} are simply Hamilton's canonical equations with ${(x_{1} , x_{3})}$ being coordinates and ${(x_{2} , x_{0})}$ their respective conjugate momenta.\footnote{The matching between ${ \{ x_{\mu} \} }$ and the canonical variables can be reordered simply by using a different choice of basis vectors in Eqs.~\eqref{eq:rotating_sigma}.}

\section{Motion around an oblate planet}
\label{sec:oblate}

As a simple demonstration, we follow H83 and apply the spinorial formalism to the secular dynamics of a satellite around an oblate planet. The averaged perturbing Hamiltonian is~\citep{beletsky}
\begin{equation}\label{eq:j2_hamiltonian}
    \mH= (C/L^3) \big[ 1- 3 \, (L_\parallel / L )^2 \big],
\end{equation}
with $C$ a constant. Here, $L$ is the satellite's angular momentum and $L_\parallel$ its projection on the planet's axis of rotation.
When expressed with the spinor components from Eqs.~\eqref{eq:spinor_LK},
they read
\begin{subequations}
\label{eq:L_oblate}
\begin{align}
\label{eq:norm_L}
    L {} & = x_0^2 + x_1^2 + x_2^2 + x_3^2 ,
\\
    L_\parallel {} & = 2 n_1 (x_0 x_2 + x_1 x_3) + 2 n_2 (x_2 x_3 - x_0 x_1) + n_3 (x_0^2 - x_1^2 - x_2^2 + x_3^2),
\end{align}
\end{subequations}
with $\bn=[n_1,n_2,n_3]$ the unit vector along the planet's rotational axis.\footnote{Note that the angular momentum in Eq.~\eqref{eq:norm_L} is a quadratic function of the spinor,
i.e., spinors are the ``square roots'' of vectors \citep{coddens17}.} After injecting the Hamiltonian from Eq.~\eqref{eq:j2_hamiltonian}, Eqs.~\eqref{eq:xdot} have an exact solution in closed form (see Appendix~\ref{app:solution}). 
Assuming $\bn=[0,0,1]$, it reads
\begin{subequations}
\label{eq:sol_oblate}
\begin{align}
    x_0 {} & = A_{+} \sin(\omega_{+} t + \delta_{+}),
    \\
    x_1 {} & = A_{-} \sin(\omega_{-} t + \delta_{-}),
    \\
    x_2 {} & = A_{-} \cos(\omega_{-} t + \delta_{-}),
    \\
    x_3 {} & = A_{+} \cos(\omega_{+} t + \delta_{+}),
\end{align}
\end{subequations}
with ${ \delta_{\pm} }$ some given phases,
the amplitudes ${ A_{\pm} }$ satisfying the constraints
\begin{subequations}
\begin{align}
     A_{+}^2 + A_{-}^2 {} & = L ,
     \\
     A_{+}^2 - A_{-}^2 {} & = L_\parallel ,
\end{align}
\end{subequations}
and the constant frequencies
\begin{equation}
    \omega_{\pm} = C \, \bigg(\! \pm \tfrac{6L^2-30L_\parallel^2}{L^6} + \tfrac{12L_\parallel}{L^5} \bigg).
\label{eq:freq}
\end{equation}
When substituting into Eqs.~\eqref{eq:spinor_LK}, we recover that both $\bL$ and $\bK$ precess with constant magnitudes~\citep{beletsky}. The critical inclination~\citep{lubowe} follows from the resonance condition $\omega_{+} = \omega_{-}$.

\section{Discussion and summary}
\label{discussion}

We followed H83 in using spinors to describe quasi-Keplerian systems on secular timescales.
In that case, an orbit-averaged Keplerian orbit is represented by a single spinor rather than two vectors.
The Poisson brackets of these spinor components turn out to be remarkably simple:
they are canonically conjugate variables. As such, the spinorial counterpart
of the vectorial Milankovitch equations is thus a set of standard canonical Hamiltonian equations.

The spinorial formalism has difficulties if the orbit is either circular or radial. In the circular case, it has an extra degree of freedom associated with the orientation of the pericentre vector, which is unphysical at circular orbits. In the radial case, the spinor is identically zero (see Eq.~\ref{eq:norm_L}), and one loses the information about the orbit's orientation. 

Future work will be devoted to testing alternative normalisations other than Eq.~\eqref{eq:norm} that could help at extreme eccentricities, as well as using the bivector formulation from Appendix~\ref{app:bivectors}.
We will also explore if the formalism above could be used efficiently in numerical integrations~\citep[see, e.g.,][]{mclachlan2014}.

\backmatter

\bmhead{Acknowledgments}
This work is partially supported by grant Segal ANR-19-CE31-0017
of the French Agence Nationale de la Recherche,
and by the Idex Sorbonne Universit\'e.

\section*{Declarations}
The authors declare no competing interests.

\noindent

\begin{appendices}

\section{The canonical basis of Delaunay variables}
\label{app:vectors}

Following the notation from~\citep{murraydermott99},
the angular momentum and pericentre vectors
can be expressed on the canonical basis of Delaunay variables via
\begin{subequations}
\begin{align}
    L_1 {} & = \sqrt{G^2 - H^2} \sin h,
    \\
    L_2 {} & = -\sqrt{G^2 - H^2} \cos h,
    \\
    L_3 {} & = H,
    \\
    K_1 {} & = G \cos g \cos h - H \sin g \sin h,
    \\
    K_2 {} & = G \cos g \sin h + H  \sin g \cos h,
    \\
    K_3 {} & = \sqrt{G^2 - H^2} \sin g ,
\end{align}
\end{subequations}
with $g$ the argument of pericentre, $G$ the magnitude of the angular momentum, $h$ the argument of node and $H$ the third component of the angular momentum vector.\footnote{These expressions follow from Eqs.~(2.119--120) in~\citep{murraydermott99}.}
 The Poisson brackets of the Delaunay variables are 
\begin{equation}
    \{ g , G \} = \{h , H \} = 1 ,
\end{equation}
while all the others are zero.

\section{Spinors and rotations}
\label{app:spinor}

Spinors are powerful tools to treat rotations. Any rotation of a vector $\bv$ can be executed as (see Eq. 2.1 in H83)
\begin{equation}
    \bv \mapsto \us^\dagger \bv \us.
\end{equation}
In order to illustrate it, let us consider the unit vector $\bsigma_1$. The spinor that rotates it by $\pi/2$ around $\bsigma_3$ is 
\begin{equation}
    \us = \tfrac{1}{\sqrt{2}} + \tfrac{1}{\sqrt{2}} \, \rI \bsigma_3.
\end{equation}
Indeed, some easy algebra using Eqs.~\eqref{eq:clifford}--\eqref{eq:clifford2} leads to
\begin{equation}
    \left( \tfrac{1}{\sqrt{2}} - \tfrac{1}{\sqrt{2}} \, \rI \bsigma_3 \right) \bsigma_1 \left( \tfrac{1}{\sqrt{2}} + \tfrac{1}{\sqrt{2}} \, \rI \bsigma_3 \right) = \bsigma_2 .
\end{equation}
This matches with the geometric intuition since the ${ \{ \bsigma_{a} \} }$
are orthogonal to one another. A general rotation by an angle $\| \btheta \|$ around an axis $\btheta$ is given by (see Eq.~{2.3} in H83)
\begin{equation}
\label{eq:general_rotation}
    \us = \sum_n^\infty\frac{\left(\frac{1}{2} \, \rI \btheta \right)^n}{n!} = \re^{\frac{\rI}{2} \btheta },
\end{equation}
where H83 normalises the spinor to unity in Eq.~{(2.2)} therein. 

H83 derives an evolution equation for the spinor (see Eqs.~{3.1}--{3.3} therein), namely
\begin{equation}
    \dot{\us} = \tfrac{1}{2} \, \rI \us \bomega,
\end{equation}
with $\bomega$ the angular velocity of the rotation (see Eq.~{3.7} in H83).
It has the simple formal solution
\begin{equation}\label{eq:general_solution_2}
    \us = \re^{\frac{\rI}{2} \bomega t} ,
\end{equation}
i.e., a general rotation as in Eq.~\eqref{eq:general_rotation}.

We finally point out that the spinors of 3D rotations are formally identical to Hamilton quaternions~\citep{hestenes83}.
Indeed, $-\rI \bsigma_1$, $-\rI \bsigma_2$ and $-\rI \bsigma_3$ obey the same algebra as Hamilton's $ \{ i, j , k \}$, respectively~\citep{hamilton1844}.

\section{Milankovitch equations with bivectors}
\label{app:bivectors}

In Sec.~\ref{sec:general} we introduced the basic concepts of geometric algebra.
We now apply it to bivectors. Similarly to vectors, we consider two arbitrary bivectors $u$ and $v$, and we define the dot and wedge product between them as the symmetric and antisymmetric part of the geometric product, namely
\begin{subequations}
\begin{align}
    u \cdot v {} & = \tfrac{1}{2} \big( uv + vu \big) ,
\\
    u \wedge v {} & = \tfrac{1}{2} \big( uv - vu \big). 
\end{align}
\end{subequations}

Let us now consider four orthonormal unit vectors ${ \bsigma_a \!\in\! \mathbb{R}^{4}} $
with $a=1,2,3,4$. We can construct a total of 6 unit bivectors out of them.\footnote{For a detailed description of the connection between the Kepler problem and four dimensional geometry, see~\citep{oliver04}.} The key point is to note that their respective wedge products
follow a structure similar to the one of the Poisson brackets
of ${ (\bL , \bD) }$, with $\bD$ the eccentricity vector~\citep[see Eq.~{(9.133)} of][]{goldstein02}.
For example, one has
\begin{equation}
    \{ L_1, L_2 \}=L_3
    \quad \Longleftrightarrow \quad
\bsigma_2 \bsigma_3 \wedge \bsigma_1 \bsigma_3 = \bsigma_1 \bsigma_2.
\end{equation}
This motivates then the correspondences
\begin{subequations}
\begin{align}
    L_1 \mapsto \bsigma_2\bsigma_3, \quad  & D_1 \mapsto \bsigma_1\bsigma_4 ,
    \\
    L_2 \mapsto \bsigma_1\bsigma_3, \quad & D_2 \mapsto \bsigma_4\bsigma_2 ,
    \\
    L_3 \mapsto \bsigma_1\bsigma_2, \quad & D_3 \mapsto \bsigma_3\bsigma_4 ,
\end{align}
\end{subequations}
Similarly, we are led to defining the bivector
\begin{equation}                            
v = L_1 \bsigma_2 \bsigma_3 + L_2 \bsigma_1 \bsigma_3 + L_3 \bsigma_1 \bsigma_2 + D_1 \bsigma_1 \bsigma_4 + D_2 \bsigma_4 \bsigma_2 + D_3 \bsigma_3 \bsigma_4 ,
\end{equation}
along with a $\nabla_{v}$ operator following the same pattern
\begin{equation}
\nabla_v = \p_{L_1} \bsigma_2 \bsigma_3 + \p_{L_2} \bsigma_1 \bsigma_3 + \p_{L_3} \bsigma_1 \bsigma_2 + \p_{D_1} \bsigma_1 \bsigma_4 + \p_{D_2} \bsigma_4 \bsigma_2 +\p_{D_3} \bsigma_3 \bsigma_4.
\end{equation}
Milankovitch Eqs.~\eqref{eq:milankovitch_vec} then become\footnote{Contrary to Eqs.~\eqref{eq:milankovitch_vec}, here we use the eccentricity vector, $\bD$, rather than the pericentre one, $\bK$.}
\begin{equation}
    \dot{v}= \nabla_{v} \mH \wedge v ,
\end{equation}
i.e., a ``rotation in bivector space''.
This equation has a first integral, ${ \rd (v \!\cdot\! v) / \rd t = 0 }$,
associated with the conserved quantity
\begin{equation}
    \big( L^2+D^2 \big) + \big( L_1 D_1 + L_2 D_2 + L_3 D_3 \big) \, \bsigma_1 \bsigma_2 \bsigma_3 \bsigma_4 = \const
\end{equation}
The two terms in this equation are conserved independently, in agreement with, e.g., Eq.~{(17)} of~\citep{tremaine2009}.

\section{Solution of the spinorial equations}
\label{app:solution}

Substituting Eq.~\eqref{eq:j2_hamiltonian}
into the spinorial equations of motion~\eqref{eq:xdot}
yields
\begin{subequations}
\label{eq:oblate_rates}
\begin{align}
    \dot{x}_0 {} & = - C \bigg[\! - \tfrac{6L^2-30L_\parallel^2}{L^6} \, x_3 - \tfrac{12L_\parallel}{L^5}(n_1 x_1+n_2 x_2 + n_3 x_3) \bigg],
    \\
    \dot{x}_1 {} & = \;\,\,\, C \bigg[\! -\tfrac{6L^2-30L_\parallel^2}{L^6} \, x_2 - \tfrac{12L_\parallel}{L^5}(n_1 x_0+n_2 x_3 - n_3 x_2) \bigg],
    \\
    \dot{x}_2 {} & = - C \bigg[\! -\tfrac{6L^2-30L_\parallel^2}{L^6} \, x_1 - \tfrac{12L_\parallel}{L^5}(n_1 x_3-n_2 x_0 - n_3 x_1) \bigg],
    \\
    \dot{x}_3 {} & = \;\,\,\, C \bigg[\! -\tfrac{6L^2-30L_\parallel^2}{L^6} \, x_0 - \tfrac{12L_\parallel}{L^5}(n_1 x_2-n_2 x_1 + n_3 x_0) \bigg] .
\end{align}
\end{subequations}

Applying Eqs.~\eqref{eq:oblate_rates} to Eqs.~\eqref{eq:L_oblate},
one gets 
\begin{subequations}
\begin{align}
    \dot{L} {} & = 0 ,
\\
    \dot{L}_\parallel {} & = 0 ,
\end{align}
\end{subequations}
i.e., both the total angular momentum and its projection on the rotation axis are conserved. Using this, Eqs.~\eqref{eq:oblate_rates} become a set of linear differential equations with an antisymmetric matrix. It results in purely imaginary eigenvalues, i.e., the solution is oscillatory.
After picking up a particular reference frame like in Sec.~\ref{sec:oblate}, the equations of motion simplify to
\begin{subequations}
\begin{align}
    \dot{x}_0 {} & =\;\,\,\, \omega_{+} \, x_3,
    \\
    \dot{x}_1 {} & =\;\,\,\, \omega_{-} \, x_2,
    \\
    \dot{x}_2 {} & =-\omega_{-} \, x_1,
    \\
    \dot{x}_3 {} & =-\omega_{+} \, x_0,
\end{align}
\end{subequations}
using the abbreviations from Eq.~\eqref{eq:freq}. These equations can be immediately integrated to give Eqs.~\eqref{eq:sol_oblate}.

\end{appendices}


\end{document}